\documentclass[12pt,preprint]{aastex}
\pdfoutput=1
\newcommand{\As}{\rm \AA}
\usepackage{graphicx}
\begin{document}
\title{
An axisymmetric hydrodynamical model for the torus wind in AGN.
~III:  Spectra from 3D radiation transfer calculations.}

\author{A. Dorodnitsyn\altaffilmark{1,2}, T. Kallman\altaffilmark{1}}

\altaffiltext{1}{Laboratory for High Energy Astrophysics, NASA Goddard Space Flight Center, Code 662, Greenbelt, MD, 20771, USA}
\altaffiltext{2}{Space Research Institute, Profsoyuznaya st., 84/32, 117997, Moscow, Russia}

\begin{abstract}
We calculate
a series of synthetic X-ray spectra from outflows originating from
the obscuring torus 
in active galactic nuclei (AGN). Such modeling includes 2.5D hydrodynamical simulations of an
X-ray excited torus wind, including  
the effects of X-ray heating, ionization, and radiation pressure. 
3D radiation transfer calculations are performed in the 3D Sobolev approximation. Synthetic X-ray line spectra and individual profiles of several strong lines are 
shown at different inclination angles, observing times, and for different characteristics of the torus.

Our calculations show that rich 
synthetic warm absorber spectra from 3D modeling 
are typically observed at a larger range of inclinations than
was previously inferred from simple analysis of the transmitted 
spectra.
In general, our results are supportive of warm absorber models    based on the hypothesis of an "X-ray excited funnel flow"  and are consistent with characteristics of such flows
inferred from observations of warm absorbers from Seyfert 1 galaxies.
\end{abstract}
\section{Introduction}

Among the most striking data taken with the grating instruments on the $Chandra$ and $XMM-Newton$ 
X-ray satellites are the rich line absorption and emission spectra
from Seyfert galaxies and other active galactic nuclei (AGN).  These spectra, referred  to as `warm absorbers', 
contain features from partially ionized ions of most astrophysically abundant elements, 
and they display systematic broadening and blueshifts corresponding
to Doppler velocities   $\sim 10^2$ -- $10^3$ km s$^{-1}$ \citep{Kasp01, Kasp02, Lee01}.

The blueshifts indicate an outflow, suggest 
mass fluxes high enough to be important to the  
AGN mass budget, and they may contain clues about the nature of the gas accreting onto
the central black hole and its influence on the surrounding galaxy \citep{Murray95}.

The
X-ray and UV spectra are rich with spectral features and of high statistical quality, and appear to be directly associated with the orientation effects inferred from Seyfert  galaxy 
unification \citep{AntonucciMiller85}.  That is, Seyfert 1 spectra show 
line absorption and scattering, while Seyfert 2s show line and bound-free continuum 
emission.  The properties of the gas, i.e. the ionization, 
speed, and column density, are similar in the two cases; the primary difference is in the 
orientation relative to our line of sight.  In Seyfert 1 objects we view the partially ionized 
gas primarily in transmission, and in Seyfert 2 objects we see it in reflection.  

Observed warm absorber outflow velocities, and virial arguments,
suggest that they are launched at $\geq 10^6$ gravitational radii, or $\sim 10^{18}$cm 
from a 10$^6$ $M_\odot$ black hole.  This coincides with estimates for the location of the material, 
the `obscuring molecular torus', which is responsible for blocking
direct views of the central black hole in Seyfert 2 galaxies
\citep{KrolikBegelman86}.  The structure of this obscuring torus, its origin and fate, are key questions in the 
study of AGN.  Although the torus may represent part of a flow originating at smaller
 \citep{Elit06} or larger  \citep{Proga07} radii,
a convenient definition is that of cool ($\sim 1000$K) optically and geometrically
thick gas in approximate rotational and virial equilibrium at $\sim$1 pc \citep{KrolikBegelman88}.
Any gas within the central region of an AGN is likely to be  heated 
and ionized by the strong UV-X-ray continuum from the black hole and inner accretion 
disk.  The temperature of X-ray heated gas asymptotically approaches the 
'Compton temperature', $T_{IC}\sim 10^7$K, when the radiation flux is 
strong enough to exceed radiative cooling driven by atomic processes
(in case of the Compton heated accretion disk winds see e.g. \citet{Bege83}). However, adiabatic losses due to the wind motion may help to keep the temperature of the gas at lower levels \citep{Dorodnitsyn08b}.

Gas at the Compton temperature will not be bound in the 
local potential and will flow out in a wind  (Krolik and Begelman 1986).  
The torus interior gas must be much cooler, and an X-ray illuminated torus will develop 
a heated intermediate region, corresponding to the transition between the 
two temperature extremes.   The details of this transition are key to the 
understanding of warm absorbers, since neither the cool interior, nor the 
hot Compton flow can produce the spectra we observe.  The structure of 
the transition depends on the density structure and, therefore, on the dynamics.
Simple estimates, based on the assumption that the gas remains in ionization 
and thermal equilibrium with the radiation field, and that the flow  
geometry is spherical, predict that the transition should be sudden, so 
there should be little gas in the intermediate `warm absorber' state.
This is clearly in conflict with the observations, suggesting that time-dependent 
or geometrical effects are important: warm 
absorbers may represent a snapshot of this gas as it is transiently 
heated and flows out from the torus.

We have constructed 2.5 dimensional (2D, axisymmetric with rotation)
time dependent models for outflows driven by X-ray heating, which incorporate the torus geometry. Simple transfer solutions suggest these 
can make spectra similar to those observed.
They produce column densities, mean ionization states, and velocities 
of the outflows which are comparable to what is inferred from observed X-ray 
spectra.   These conclusions are backed up by detailed multi-dimensional numerical 
hydrodynamic calculations, incorporating the physics of radiative heating, cooling, 
photoionization, and radiation pressure \citep{Dorodnitsyn08a, Dorodnitsyn08b}.    A challenge for these  models is that, 
since the warm absorber gas we observe is in a transient state, it occupies 
only a fraction of the volume in the torus opening.  This predicts detectable
warm absorbers from a smaller fraction of  AGN than is observed, since for
many inclination angles the line of sight to the X-ray continuum source 
misses the partially ionized gas. 
The observed fraction is $\sim 50 \%$, and perhaps greater (Reynolds, 1997), while models 
predict approximately half this number.  Furthermore, the models calculated 
so far do not reproduce some of the features observed from X-ray spectra, 
including an apparent bimodal distribution of ionization in many objects, 
and the presence of X-ray absorption by dust or neutral material in some spectra.

Some of the conclusions of our past work are limited by the simple 1-dimensional transfer we employed to calculate the emergent spectra.
Although these were useful as a preliminary step of predicting line spectrum, 
the 1D radiation transfer calculations are not capable of producing emission lines and of treating effects like filling of absorption lines with emission lines. In addition, all effects associated with non-radial motion of the fluid were neglected.

In this paper we produce full 3D radiation transfer calculations of X-ray line spectra from 
such flows using the information provided by our hydrodynamical models \citep{Dorodnitsyn08a, Dorodnitsyn08b}. These simulations produce spatial distributions of the projected velocities, and densities. Using this information as an input to the 3D radiation transfer in X-ray lines, we perform 3D radiation transfer in the Sobolev approximation. We calculate spectra in approximately $10^5$ lines and also high resolution profiles of some of the strongest lines. 

\section{Methods}
To calculate an X-ray line spectrum from the moving gas which was stripped off the AGN torus via X-ray induced evaporation, we combine together the following ingredients:

i) Taking into account heating and cooling processes we calculate a set of time-dependent 2.5D hydrodynamical models of winds evaporated from the AGN torus. 
ii) Based on the 3D generalization of the escape probability method, we construct a numerical code which is able to calculate line spectra from a 3D flow.
iii) Making use of the XSTAR code we calculate tables of X-ray opacities for different energies and ionization parameters.
iv) Using opacity tables and distributions of density, and velocity from hydrodynamical models, we calculate the emergent spectrum. Step i) has been described in \cite{Dorodnitsyn08a}. In the following we briefly review methods and describe results of steps ii) through iv).

\subsection{Hydrodynamical modeling}

Our hydrodynamic and radiation framework is described in detail
\cite{Dorodnitsyn08a, Dorodnitsyn08b}. Here we briefly review the methods and results. 

The primary physical assumption of our model is that
that the inner face of the rotationally supported torus is exposed to radiation heating by X-rays generated by accretion near the supermassive black hole. 

The equations of hydrodynamics are 
calculated using the ZEUS-2D code \citep{StoneNorman92}, modified to take into account various processes of radiation-matter interaction; this modeling includes time-dependent 2.5D hydrodynamical simulations (that is time-dependent 2D simulations with rotation) of the flow. 
Interaction of the radiation with matter is incorporated into the hydrodynamical code taking into account radiation heating and cooling in the energy equation and pressure of the radiation in UV spectral lines and in  continuum in the momentum equation.
The rates of Compton and photo-ionization heating and Compton, radiative recombination, bremsstrahlung and line cooling are calculated using approximate formulae, which are modified from those of \cite{Blondin94}  \citep{Dorodnitsyn08b}. 
These modified formulae adopt newly available atomic data and heating and cooling rates obtained from the XSTAR code \citep{KallmanBautista01}.  

We assume that a certain fraction of a total black hole (BH) accretion luminosity, is emitted as X-rays which
interact with the inner face of the rotationally supported torus and heat it. Thus the cold gas of the torus is heated to almost the Compton temperature ($T\sim10^7$), fills up a throat of the torus constituting an outflow. This hypothesis was tested both qualitatively and quantitatively by evolving such models in time, and exploring the parameter space, including the accretion efficiency, and geometrical and physical properties of the torus.

\subsection{Radiation transfer in spectral lines}

In the papers by \cite{Dorodnitsyn08a, Dorodnitsyn08b} pure absorption spectra were calculated from different models assuming different inclinations. Transfer was calculated using simple 1D attenuation of the radiation due to line and bound-free absorption, assuming pure radial streaming of the radiation and also approximating the emitting core as a point source. The obvious limitations of this approach motivates a much more realistic radiation transfer treatment which would take into account spatial gradients of the velocity in a 3D flow.
In this work we implement such an approach, in a 3D radiation transfer calculations in Sobolev approximation for the line radiation transfer.

The supersonic motion of gas of the AGN wind provides justification for use of the Sobolev approximation while calculating radiation transfer in spectral lines.
The problem of radiation transfer in spectral lines in the Sobolev approximation in a spherically-symmetrical flow is addressed in numerous papers (for references see e.g. 
\cite{RybickiHummer78,RybickiHummer83, Dorodnitsyn09} ). 
The primary assumption of the Sobolev approximation is that 
due to velocity gradients in the flow the Doppler shifting of frequency moves a photon quickly out of the resonance soon after it interacts with the flow. Thus the radiation transfer in the Sobolev approximation is treated as an entirely local problem, assuming the characteristics of the absorbing plasma are constant across the narrow region of interaction.
Our  hydrodynamical model is axially symmetric, but this symmetry is lost if viewed from an inclination angle, so
to calculate the spectrum
we must adopt a 3D Sobolev approximation. 
Even if the poloidal velocity along the streamline is monotonically increasing, the {\it projected velocity} may not behave in such a way, rather having a maximum or several maxima (depending on the real structure of the 2D flow). This poses the problem of multiple resonances which we discuss in detail below.

\begin{figure}[htp]
\includegraphics[width=260pt]{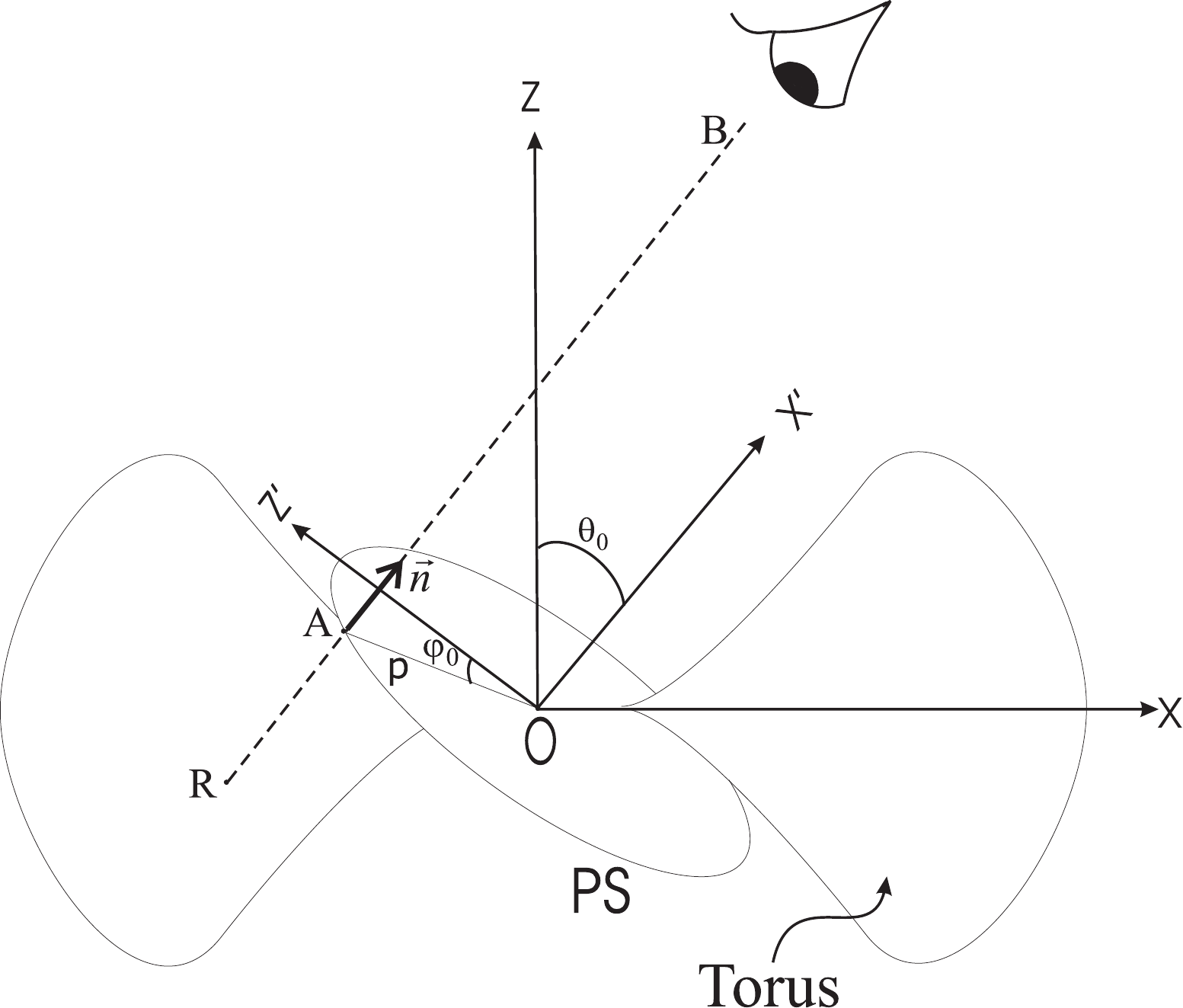}
\caption{Illustration of the geometry and coordinate systems implied by the calculations (see text for details). Notice, that the $y$-axis (not shown) points into the plane of the picture to form a right-handed $xyz$ coordinate system, and  $y'=y$. Not to scale. 
}
\label{Fig_geom}
\end{figure}

The geometry of the problem is shown in Fig.\ref{Fig_geom}.
The Cartesian coordinate system $xyz$ ($y$ axis is not shown) is located at $O$. The observer is situated at infinity looking at the torus from an inclination, $\theta_{0}$. Since the hydrodynamical model is axially-symmetric, 
we put the observer in the $xz$ plane. The plane of the sky, $PS$, is transported to $O$, providing another Cartesian coordinate system $x'y'z'$, ($y'$ axis is not shown) which is the $xyz$ system, rotated by the angle $\theta_{0}$ around the $y$ axis. Notice also that $y'=y$. Our assumptions imply (see further in the text) that a photon after being scattered in a spectral line, for example, in the point $R$, propagates along the straight line $RB$ in the direction of the unit vector $\bf n$.
This ray, $RB$ intersects the plane $PS$ at the point $A$, being at the angle $\phi_{0}$ from the $OZ'$ axis and at a distance $p$ (impact parameter) from $O$.
Summing all such possible rays, i.e. integrating over the plane, $PS$, the radiation flux which is observed at infinity is obtained.

The total radiation flux which is registered by the observer at infinity at the frequency, $\nu^{\infty}$, equals to

\begin{equation}\label{flux_as_int_atinf}
F( \nu^\infty, \theta_{0} ) = 
\int_0^{ 2\pi} \int_0^{\infty}\,I^{\infty} (\nu^\infty, \theta_{0}, \phi_{\rm 0}, p)\,p\,dp\,d\phi_{0}
\mbox{,}
\end{equation}
where $F_c$ is the flux in the continuum near the line frequency.

In our treatment of emission only resonant lines are considered. Although we do consider non-resonant absorption in the approximate treatment of the attenuation (section 2.3).
After being emitted by the core, or, alternatively, being conservatively scattered in a line, a photon remains at constant frequency, $\nu^{\infty}$ and direction of propagation, unless 
its frequency, $\nu_{\bf com}$ in the frame co-moving with the gas appears in a resonance with a line transition:

\begin{equation}\label{nucomoving}
\frac{|\nu_{\rm com} - \nu_{l}|}{\Delta\nu_{D}}= \frac{1}{\Delta\nu_{D}}\left(\nu^{\infty}(1-\frac {\bf v \cdot\, n}{c})
- \nu_{l}\right) \lesssim 1\mbox{,}
\end{equation}
where $\nu_{l}$ is the frequency of the line, and $\Delta\nu_{D}$ is the Doppler width of the line. For such a $\nu^{\infty}$, all such possible resonant points constitute resonant surfaces, which could have an extremely complicated shape in the case of a real flow. After encountering the resonance surface the intensity $I$ changes discontinuously:

\begin{equation}\label{Intens_jump_onEFS}
I^{s_i+0}=I^{s_i-0}e^{-\tau_{l,i}}+
S_i(1-e^{-\tau_{l,i}})=I_{\rm dist, i}+I_{\rm loc, i}\mbox{.}
\end{equation}
where
$S_{i}=S (\sigma_i)$ is the source function and $\tau_{l,i}=\tau (\sigma_i)$ is the optical depth in a spectral line. 
The intensity changes discontinuously, being a) attenuated by a  
$e^{-\tau_{l,i}(\sigma_i)}$ factor, and b) reinforced by the contribution of the locally scattered radiation field,
$I_{\rm loc,i}=S_i (1-e^{-\tau_{l,i}}) \sim 1/r^2 
(1-e^{-\tau_{l,i}})$ (the last equality holds true for a point source and pure line scattering). The latter contribution results from scattering of external photons at a resonant point.
Let $s$ be the distance measured along the photon's trajectory, then $\tau_{l}$ in a direction ${\bf \hat n}$ in a moving gas depends on the gradient of the velocity in this direction, i.e. on $dv/ds$ and can be cast in the form:

\begin{equation}\label{oticalDepth}
\tau_{l}=\frac{\rho\,\epsilon \,\kappa_{es} v_{th}}{|dv/ds|}\mbox{,}
\end{equation}
where $\epsilon=\kappa_{l}/\kappa_{es}$, $\kappa_{l}$ is the line opacity, $\kappa_{es}$ is the Thomson scattering opacity, $\rho$ is the density, and
$v_{th}$ is the thermal velocity.
 
In the presence of multiple resonances, distant points within the flow are causally connected, i.e. they can exchange photons ranging within the line frequency width. 
In terms of surfaces of equal frequencies, such surfaces become multi branched, allowing for a photon of a frequency $\nu^\infty$ to interact with the moving matter at different points along a single line of sight.
In the presence of multiple resonance surfaces, we assume that the interaction between $\sigma_{i-1}$ and $\sigma_{i}$
takes place only in the forward  direction (in the direction towards the observer). All other possible interactions are neglected. 
Then the expression (\ref{Intens_jump_onEFS}) should be just summed over the available resonance surfaces.
Altogether, these assumptions are known as the {\it disconnected approximation} \citep{GrachevGrinin,Marti77,RybickiHummer78,RybickiHummer83}:

\begin{equation}\label{IntensityInfinity}
I^{\infty}(\nu^{\infty})= I^{\rm ins , 0}\,e^{-\sum_i^N\, \tau_{l i} }
+\sum_{i}^{N}\, S_{i} \left(1-e^{-\tau_{l i}} \right)\,e^{-\sum_j ^{N-1}\,\tau_{l j}} 
\mbox{,}
\end{equation}
where $N$ is the total number of resonances and $I^{\rm ins , 0}=I_{\rm src}\, e^{-
\tau_{es} ( {\bf r_{0} } )}$ is the initial intensity of the radiation at the point 
${\bf r}_{0}$, which is the farthest point along the ray from the  observer in the computational domain, $I_{\rm src}$ is the intensity emitted by the source (core), and for simplicity, we omit the dependence of $I$ on $p$, $\phi_{0}$, and $\theta_{0}$; $\tau_{es}$ is an electron scattering optical depth: 

\begin{equation}\label{Thomson_tau}
\tau_{es}({\bf r})=\int_0^{r}\kappa_{es}\rho\,dr \mbox{.}
\end{equation}

The
source function is calculated with the method of escape probabilities (e.g. \citet{Castor}).
The probability of a photon to escape in a direction ${\bf \hat n}$ in a moving gas, depends on the gradient of the velocity in this direction, ${\bf \hat n}$, i.e. on $dv/ds$:

\begin{equation}
\frac{dv}{ds}=\sum_{ij}\,e_{i,j}\,n_{i}\, n_{j}\mbox{,}
\end{equation}
where $e_{i,j}$ is the rate of strain tensor:

\begin{equation}\label{eij}
e_{i,j}=\frac{1}{2}\left(\frac{\partial v_{i}}{\partial r_{j}}+\frac{\partial v_{j}}{\partial r_{i}}\right)\mbox{,}
\end{equation}
Components of (\ref{eij}) in spherical polar coordinates, are given, for example, by  \cite{Batchelor}.

A probability, $\beta_{\bf n}$ of a photon to escape in a direction 
$\bf n$ is given by

\begin{equation}
\beta_{\bf{n}} = (1-e^{-\tau( {\bf n} )} )/e^{-\tau( {\bf n} ) }
\mbox{.}
\end{equation}
Integrating this relation over solid angles $\Omega({\bf n})$ (i.e. either all over the sky, or over solid angles subtended by the source, $\Omega_{\rm src}$)  
results, respectively, in escape and penetration probabilities \citep{RybickiHummer83}:

\begin{equation}
\beta_{\rm esc}=\frac{1}{4\pi}\int_{4\pi} \,\beta_{\bf{n}}\,d\Omega
\mbox{.}\label{escProb}
\end{equation}

\begin{equation}
\beta_{\rm pen}=\frac{1}{4\pi}\int_{\Omega_{\rm src}}\,\beta_{\bf{n}}\,d\Omega
\mbox{.}\label{escProb_pen}
\end{equation}
In the case of pure line scattering, the source function is cast in the form:

\begin{equation}\label{SourceFunction}
S({\bf r})=\frac{J({\bf r})}{\beta_{\rm esc}}\mbox{,}
\end{equation}
where $J$ is the mean intensity at the point $\bf r$:

\begin{equation}\label{Jc1}
J({\bf r})=\frac{1}{4\pi} \int_{\Omega_{\rm src}}\,\beta_{\rm n}{\tilde I}_{\rm src}({\bf r}) \, d\Omega\mbox{,}
\end{equation}
and ${\tilde I}_{\rm src} ( {\bf r} )$  is the intensity of the continuum near 
$\nu_{l}$, possibly attenuated by the electron scattering prior to resonant point, ${\bf r}$:
${\tilde I}_{\rm src}({\bf r})=I_{\rm src}\,e^{-\tau_{es}}$. It is assumed that the attenuation is dominated by Thomson scattering
 $\kappa_{es}=0.2(1+X_{\rm H})\simeq0.4\,{\rm cm^{2}\,g^{-1}}$, where $X_{\rm H}$ is the mass fraction of hydrogen,  and the factor $e^{-\tau_{es}}$,
accounts approximately for the attenuation of the radiation flux on the way from the source to the resonant point. 

Pre-calculated hydrodynamical models provide the distributions of density, $\rho({\bf r})$ 
and velocity, ${\bf v}({\bf r})$, needed to calculate $S_{i}$ and $\tau_{l i}$. With this knowledge in hand, substituting $I^{\infty}(\nu^{\infty})$ from  (\ref{IntensityInfinity}) to (\ref{flux_as_int_atinf}), we calculate the
spectrum. Notice that in case of spherically symmetrical distributions of the velocity, density, etc., the above formalism gives the well known P-Cygni profiles. 

We have rigorously tested our calculations against the atlas of P-Cygni profiles of \cite{CastorLamers}. Results of such testing are shown in 
\cite{Dorodnitsyn09}, Figure 9. 

\subsection{Treatment of lines and continua}

First we point out that in this work we consider transitions in {\it all} lines as resonant. That is, we don't differentiate between true resonant event and lines whose upper levels are depopulated by photoionization, auto-ionization, and thermalization processes. 
This is an approximation which has only a small effect on our final results, since the strongest features in our spectra are true resonance lines.
Using the XSTAR code we calculate two opacity tables:
table 1 contains only line opacities, and table 2, only
continua opacity. Both tables contain cross sections as functions of ionization parameter (see further in the text) and energy.  As mentioned in section 2.2, to integrate flux, $F$, from (\ref{flux_as_int_atinf})
we sum all rays with all impact parameters, $p$ at all $0<\phi< 2\pi$ (see Figure \ref{Fig_geom}). Starting from the boundary of the computational domain (which is at $r=50$pc), we calculate intensity along the ray RB.
At each point we calculate the local Thomson optical depth between this point and the nucleus (located at r=0.05pc), then calculate the attenuated ionization parameter,  and then refer to table 1 to check the resonant condition, (\ref{nucomoving}), and if succeeded calculate the new value of intensity, $I$, using (\ref{Intens_jump_onEFS}). Between resonant points we attenuate the radiation field, making use of pure continuum opacities, from table 2.  We do not include any turbulent broadening in any of our calculations, rather just thermal broadening plus the effects of the bulk flow via the Sobolev optical depth.

\subsection{Assumptions and input parameters}
Our hydrodynamical modeling is based on the solution of the 2D, time-dependent hydrodynamical equations for rotating flow exposed to external X-ray heating. Radiation pressure is taken into account both in continuum and due to UV pressure in spectral lines in which case it is calculated from the generalization of the \cite{CAK75} formalism \citep{Dorodnitsyn08a, Dorodnitsyn08b}. 

The assumed photo-ionization
equilibrium of the wind implies the outflowing gas 
can be parameterized in terms of the ratio of radiation energy density to baryon density \citep{Tarter69}. One popular form of this parameter is 

\begin{equation}\label{smallxi}
\xi=4\,\pi\,F_{\rm x}/n\mbox{,}
\end{equation}
where $F_{\rm x}=L_{\rm x}e^{-\tau}/(4\pi r^2)$ is the local X-ray  flux, $L_{\rm x}$ is the X-ray luminosity of the nucleus, and  and $n$ is the number density. 
We assume that the nucleus or the ''core'' at radius 0.05 pc, emits a continuum power law spectrum with energy index $\alpha=1$.

The ionization parameter $\xi$ can be estimated in the following approximate way:
$\xi\simeq 4\cdot 10^2 \cdot f_{\rm x}\,\Gamma\, M_6/ (N_{23}\,r_{\rm pc})$, where 
$N_{23}$ is the column density in $10^{23}$ ${\rm cm}^{-2}$, $f_{\rm x}$ is a fraction of the total accretion luminosity $L_{\rm BH}$ available in X-rays,  $\Gamma$ is a fraction of the total 
Eddington luminosity $L_{edd}=1.25\cdot 10^{44}\,M_6$, where $M_{6}$ is the mass of a BH in $10^{6}M_{\odot}$. The radiation pressure in lines is calculated assuming the relative fraction of UV radiation, $f_{\rm UV}$ =0.5.
Heating and cooling rates account for
Compton and photo-ionization heating and Compton, radiative recombination, bremsstrahlung and line cooling, and are incorporated into the equations of hydrodynamics. 

After non-dimensionalisation, our problem includes several characteristic scales: the radius is measured in terms of  $R_{0}$, which is the distance of the initial torus density maximum from the BH; time in terms of 
$t_0=R_0^{3/2}/\sqrt{GM}\simeq 4.5\cdot10^{11}\,r_{\rm pc}^{3/2}\,M_6^{-1/2}\,({\rm s})$,
where $r_{\rm pc}$ is the distance in parsecs; the characteristic velocity reads
$V_0=\sqrt{GM/R_0}\simeq 6.6\cdot 10^6 M_6^{1/2} \, r_{\rm pc}^{-1/2}\,({\rm cm\,s^{-1} })$ . From now on we measure time, $t$ in units of $t_{0}$.

Initial conditions at $t=0$ are provided by the equilibrium toroidal configuration in which a polytropic, rotating gas is in equilibrium with the force of gravity and radiation force.
Thus, the initial state is a rotationally supported torus of \cite{PP84}, which is corrected for the radiation pressure.
The parameter that determines the relative distortion of such a torus is
$d=(\varpi^-+\varpi^+)/2$, where $\varpi$ is the cylindrical radius, and the inner and outer edges of the torus are located at $\varpi^-$ and $\varpi^+$, respectively.

\section{Results}
The major goal of this paper is to calculate emission and absorption spectra predicted by hydrodynamical simulations. 
The maximum density of the initial torus, $n_{\rm max}$, or, alternatively, its initial Compton optical depth $ \tau_\bot^{\rm C}=\tau( \theta=90^\circ )$
is used as a parameter to distinguish between different models;
$n_{\rm max}$ roughly scales with the column density of the absorbing gas during the evolution of the torus.

Our modeling of the spectra use a set of pre-calculated hydrodynamical
models, including combinations for: $\tau_\bot^{\rm C}=1.3 \,({\rm models\, A_i})$, and
$\tau_\bot^{\rm C}=40$ (models $B_i$); equivalently, these correspond to initial tori, having  $n_{\rm max}=10^{7}{\rm \,cm^{-3}}$ or $n_{\rm max}=10^{8}\,{\rm cm^{-3}}$. Other parameters, determining the torus are$R_0$ and the distortion parameter, $d$. Models $A_i$ and $B_i$ have $d=2.5$
Additionally there are two models with
$d=5$ (models $C_i$) which represent an extreme case of extended and rarified torus. The full list of models is given in  Table 1 of \cite{Dorodnitsyn08b}.

Spectra calculated from hydrodynamics contain too many features for easy interpretation, and to make a comparison with observations feasible we convolve them with a Gaussian which has FWHM(E)$=10^{-3}E$; 
such convolved spectra are most closely related to what would be observed by an instrument such the {\it Chandra} grating. However, contrary to such a real observer, we are also aware of un-convolved spectra, and we frequently use these to compare and interpret the results of our analysis and to establish possible pitfalls.
Further, we show that the interpretation of individual lines from the convolved spectra may contain significant caveats leading to a misinterpretation of the positions of some lines and of their relative blue-, and redshifts. 
Our spectra contain absorption as well as emission features and for 
convenience we denote the wavelength (or, equivalently the energy) of an absorption line as $\check{\lambda}$ and of an emission line as $\hat{\lambda}$;  the line centroid as $\lambda_{0}$; the blue-ward edge as $\lambda^{+}$, and the red-ward as $\lambda^{-}$.
In this paper when speaking about individual lines we adopt the following convention: velocities corresponding to blueshifts are positive and  to redshifts are respectively negative.

From hydrodynamical modeling \citep{Dorodnitsyn08a}, it is established that by the time $t\sim 3$ the torus is losing mass via a quasi-stationary evaporative wind. Again, times are measured in terms of $t_{0}$. Before this time, say at $t\sim 1$ the influence of the initial switching on of the X-ray heating is felt by the wind + torus system, and at much later time, say at $t\sim 6$, the torus becomes significantly influenced by the process of evaporation.
Thus for our investigation we choose evolution stages at which the torus possesses a well developed wind but still is not drastically influenced by it. The
results of our modeling for models $B_{6}$, $B_{3}$, and $A_{3}$ of \cite{Dorodnitsyn08b} at t=4 are shown
in Figures \ref{FigB6_T4} - \ref{FigA3_T4}.


\subsubsection*{Results for the $B_{6}$ model} 
\begin{figure}[htp]
\includegraphics[width=520pt]{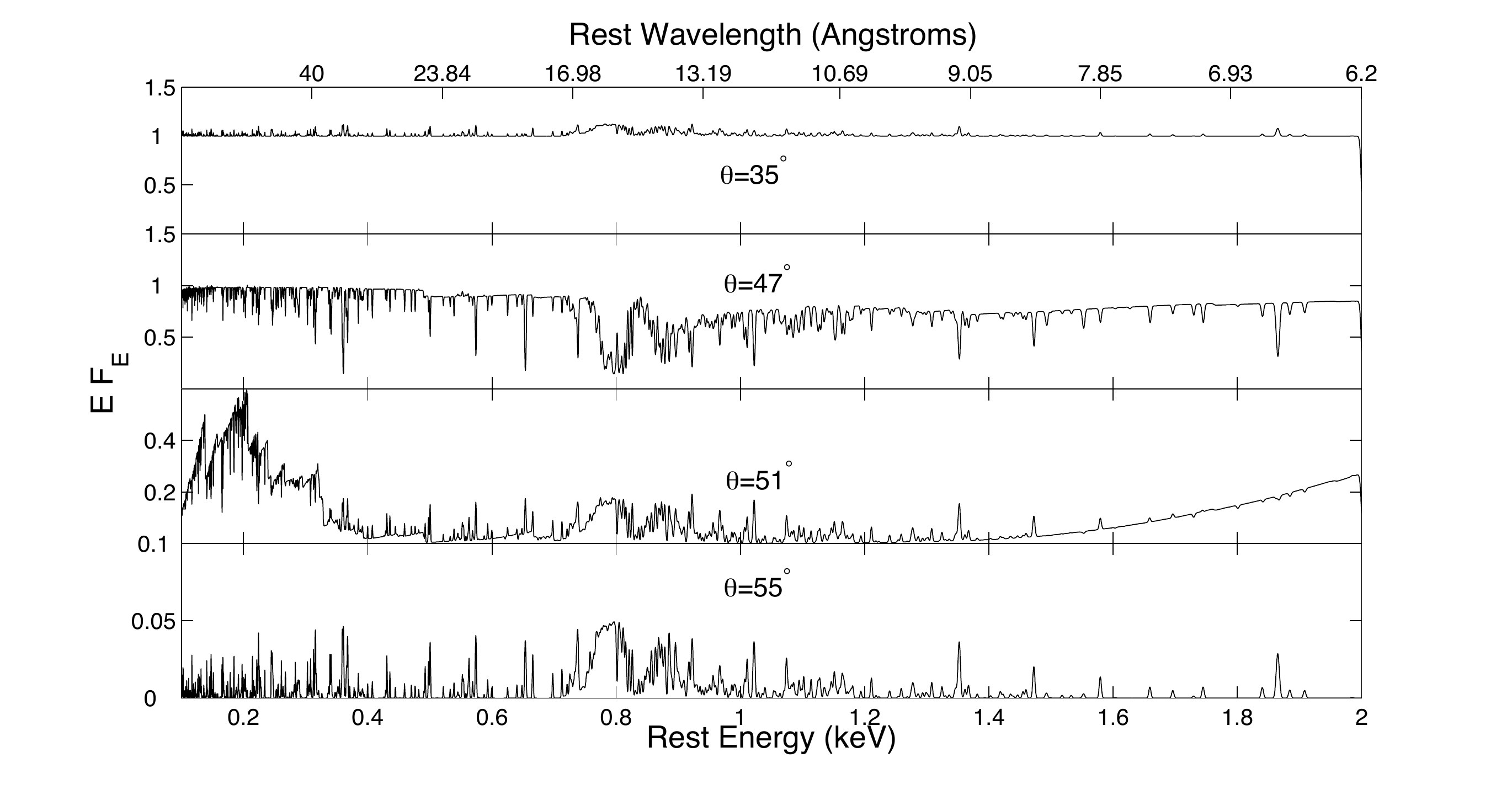}
\caption{
Model $B_6$ ($\tau_\bot^{\rm C}=40$, $R_0=1{\rm pc}$) X-ray spectra,
observed at $t=4$ as a function of angle.
}
\label{FigB6_T4}
\end{figure}

Results for this model at t=4 are shown in Figure \ref{FigB6_T4}.
The initial state has $R_{0}=1\,{\rm pc}$, $\tau_\bot^{\rm C}=40$, $n_{\rm max}=10^{7}{\rm \,cm^{-3}}$, and $d=2.5$.

Characteristics of the individual lines are consistent with hydrodynamical models.
At $\theta\simeq35^{\circ}$ mostly emission lines are seen atop of the continuum spectrum and 
at $\theta = 51^{\circ}$, emission lines are observed atop a significantly absorbed continuum. 
The position of a blueshifted $\rm O\,{VIII}\,{\rm K}\alpha$, 18.97\As  (653.5\,{\rm eV}) line remains approximately constant at all inclinations. For example, at $\theta = 45^{\circ}$ the velocity of the gas is $-725<v<880$ km ${\rm s^{-1}}$;  the line centroid is $\check{\lambda}_{0}=18.96\As$ ($653.7$) eV, corresponding to $v\simeq$ 55 km ${\rm s^{-1}}$, which is comparable with the escape velocity at $1\,{\rm pc}$.

\begin{figure}[htp]
\includegraphics[width=500pt]{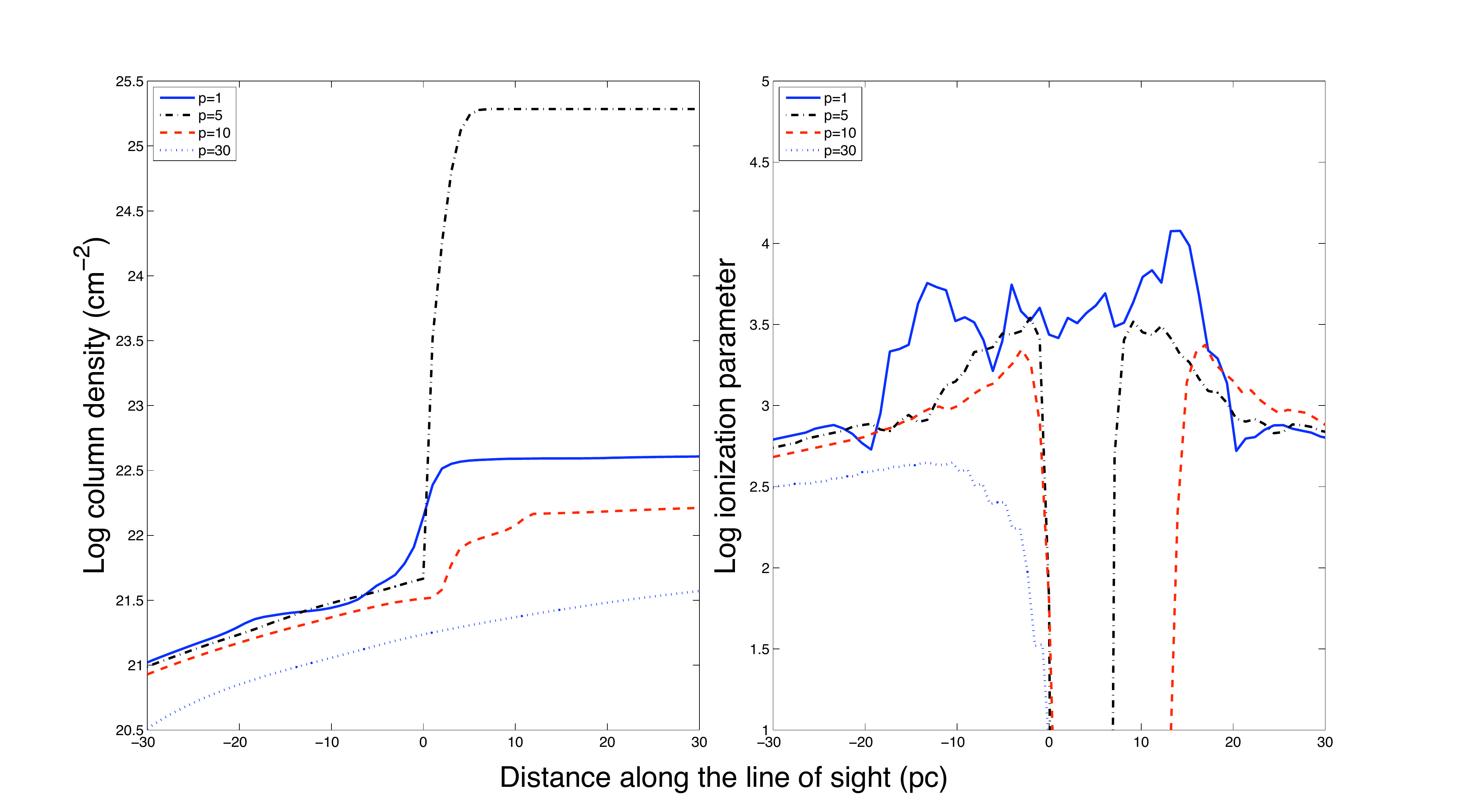}
\caption{
Model $B_6$,
observed at time $t=4$ and at inclination $\theta_{0}=45^{\circ}$ as a function of an impact parameter, $p$; the angular coordinate in the plane of the sky, $\phi_{0}=45^{\circ}$. Left panel: column density; right panel: ionization parameter. Legend: each curve is calculated for the impact parameter, p, measured in parsecs.
}
\label{FigB6_T4LineSIght}
\end{figure}

At $\theta = 47^{\circ}$ a strong absorption line is observed at $\check{\lambda}_{0}=16.8\As$   (738.3) eV in the convolved spectrum. If this line is due to ${\rm O\,{VII}}$, at  $17.2{\rm \AA}$ $(720.84{\,\rm eV}$ ) this transition indicates an $7198\, {\rm km \, s^{-1}}$ outflow.  The escape velocity at distances which are plausible for the formation of such absorption features ($\sim 1\,\rm pc$) is significantly smaller, $\lesssim 50-60 \, {\rm km\, s^{-1}}$, implying such an interpretation is unlikely. In fact, looking at the un-convolved, high resolution spectrum reveals a complicated picture in this energy range 
$\sim 17.2 - 16.7\, \As\,
(720 - 740 \,{\rm eV}$), consisting of many absorption lines which are blended together in the convolved spectrum. From the un-convolved spectrum we attribute a $\check{\lambda}_{0}=15.2\As \, (722.1{\,\rm eV} )$ feature in the un-convolved spectrum 
to ${\rm O\,{VII}}, 17.2\As \, (720.84{\,\rm eV})$ line, blueshifted with
$v \sim 524 \,{\rm km\, s^{-1}}$. The position of the ${\rm O\,{VII}}, 16.8\As \, (739.3{\,\rm eV})$ edge, clearly seen in the un-convolved spectra indicates $v \sim 700 \,{\rm km\, s^{-1}}$. This is indicative of a strong variation of the ionization parameter, $\xi$ along the line of sight in the spectrum forming region (compare to Figure \ref{FigB6_T4LineSIght} or to Figure 4 from \citet{Dorodnitsyn08b}).

The attenuation of the radiation field at high inclination is consistent with increasing column density, $N({\theta})$ at high $\theta$.
The {\it radial} column density rises from $N_{22}\simeq 1$ at $r=0.93\,{\rm pc}$ to 
$N_{22}=3.3$ at $r=2.9\,{\rm pc}$ up to $N_{22}=75$ at $r=3.9\,{\rm pc}$, where $N_{22}$ is the column density in terms of $10^{22} \, {\rm cm^{-2}}$. 

The line centroid, $\check{\lambda}_{0}$ is determined by two regions, where the absorption is most favorable: starting, approximately from $r=0.8\,{\rm pc}$, where $v_{r}\simeq v_{\phi} \sim 115 \,{\rm km\, s^{-1}}$ and density is $n\simeq 4\cdot 10^{3}{\rm cm^{-3}}$; at $r=1.8\,{\rm pc}$ the density, n slightly drops, but the velocity rises: $v_{r}\simeq 300 \,{\rm km\, s^{-1}}$; at
$r=3.56\,{\rm pc}$ the density has a maximum on the line-of-sight: $n\simeq 2.7\cdot 10^{5}{\rm cm^{-3}}$, but the velocity is only $v_{r}\simeq 26 \,{\rm km\, s^{-1}}$. The ionization parameter, $\xi$ has a plateau, where $\xi\simeq 700$, then it behaves like inverted number density, i.e. drops to zero at $r\simeq3.5\,{\rm pc}$, where $n$ in turn has a maximum.

An apparent emission hump at $\sim 14.9 -17.2\As,\ (720 - 830$ eV), $\theta=47^{\circ}$ is due to H-, and He-like oxygen and moderately ionizated iron, ${\rm Fe\,{XVII} - Fe\,{XXIV}} $. It has an onset near the ${\rm O\,{VII}}$ edge at 16.8$\As$ (739 eV) and is bounded by the
${\rm O\,{VIII}}$ edge at 14.3$\As$ (870 eV) from the high energy side.
 At higher inclinations this emission excess is substituted by the absorption trough seen roughly at the same frequencies. 
As shown in Figure 4 of \cite{Dorodnitsyn08b} it is approximately this inclination where there is a significant rise in column density and the line of sight goes though the dense and low ionization gas. 

From the strong absorption line of ${\rm Mg\,{XI} }$ at $\theta=47^{\circ}$ at $\check{\lambda}_{0}=9.2\As$ (1353 eV),  we infer $v\sim 221\, {\rm km \, s^{-1}}$; from the high resolution spectra we get $-665<v<887$ km ${\rm s^{-1}}$. At smaller inclinations this line is not prominent; the 
$\rm Mg\,XII$ line $8.4\As$ (1472 eV) is most prominent at $\theta=47^{\circ}$ giving $\check{\lambda}_{0}= 8.4\As \,(1473{\,\rm eV})$, $-407<v<611$ km ${\rm s^{-1}}$, and also from the line centroid: $v \sim 203 \,{\rm km\, s^{-1}}$.

The emission spectrum contains many lines.
For example, emission is seen close to oxygen edge, e.g. ${\rm O\,{VII} }$, indicating 
$v \sim 162 \,{\rm km\, s^{-1}}$. At slightly lower energies many lines are blended together, producing a feature $\sim 0.3\As $ (10 eV) wide. 
Among other features, most prominent are those from ${\rm Mg\,{XI} }$, ${\rm Mg\,{XII} }$, ${\rm Ne\,{X} }$. Blueshifts of these lines are consistent with a wind which is moving with velocity, $ 50 \lesssim v  \lesssim 600 \,{\rm km\, s^{-1}}$. 
Positions of emission lines, i.e. $\hat \lambda^{-}$, $\hat \lambda_{0}$, and $\hat \lambda^{+}$ are consistent with that inferred directly from the hydrodynamical models (i.e. for example with those given in Figure 4 of \cite{Dorodnitsyn08b}).

At  $\theta=10^{\circ}$, the maximum projected velocity is determined by the maximum $v_{r} \simeq 664 \,{\rm km\, s^{-1}}$ at $r\simeq 3.3\,{\rm pc}$. The toroidal velocity, $v_{\phi}$ peaks at $r\simeq 0.37\,{\rm pc}$ at a value $v_{\phi}\simeq 500 \,{\rm km\, s^{-1}}$. The column density smoothly rises from $N_{22}=0.15$ at $r \simeq1\,{\rm pc}$ to $N_{22} \simeq 1.8$ at $r=20\,{\rm pc}$.

Comparison of the spectra given in Figure \ref{FigB6_T4} with those given in Figure 10 of \cite{Dorodnitsyn08b} (which shows the pure transmitted spectra for the same model $B_{6}$) demonstrates significant differences. 
That is, the full 3d transfer models allow for more transmitted flux at a given viewing angle.  
These differences are partially due to the higher energy resolution of the opacity tables adopted in the current simulations and partially because of the qualitative difference of the 3D approach (presence of significant emission component, finite size of the core).

\subsubsection*{Obscuration of the emission lines}
No significant negative velocities are detected in emission lines. This is the result of a severe obscuration at high impact parameters, $p$.
Let $L$ be the distance between points $A$, and $R$, Figure  \ref{Fig_geom}.
In Figure \ref{FigB6_T4LineSIght} is shown the behavior of  the column density,
$N(L,\theta_{0},\phi_{0})$ and the ionization parameter, $\xi(L,\theta_{0},\phi_{0})$ with distance $L$. Figure \ref{FigB6_T4LineSIght} shows these quantities at $\theta=45^{\circ}$, $\phi_{0}=45^{\circ}$. 
The overall behavior indicates a significant absorption at high $p$. 
What we see is the line-of-sight penetrating more cold gas as it moves from $p=1$ to $p=5$. Then, at larger $p=10$ there is again less gas (this is  the end of a torus), and reducing to smaller values at $p=30$. 
Note that if there is only resonant scattering in lines present then significant emission can be obtained by accumulating photons form a large volume, which has a small Thomson depth.
Our result shows that it is difficult to see extended regions behind the torus, and thus no significant emission lines with negative velocities are seen in the simulations. 

\subsubsection*{Results for the $B_{3}$ model} 
This model represents a wind which is launched closer to the BH, at $R_{0}=0.5$ pc. The overall spectra are similar to that of $B_{6}$ model although the larger escape velocity, $V_{\rm esc}(R_{0})$ implies larger velocities inferred from individual lines. 

The prominent
$\rm O\,{VIII}\,{\rm K}\alpha$ line is blueshifted:
$ \check{\lambda}_{0} = 19\As$ (653.8 eV), 
$v\simeq$ 100	km ${\rm s^{-1}}$ at $\theta=45^{\circ}$. At higher inclinations,  $\theta=47^{\circ}$, it is stronger but remains with the same blueshift; at $\theta=51^{\circ}$, $ \check{\lambda}_{0} \simeq 19\As$. Such blueshift is consistent with the gas moving at $V_{\rm esc}(1\rm  pc)$.
At higher energies, among strong absorption lines are those of 
${\rm Ne\,{X} }$, $v\sim 500{\,\rm km\, s^{-1}}$, 
${\rm Mg\,{XI} }$ and ${\rm Mg\,{XII} }$, indicating $ |v| \lesssim 1000\,  {\rm km\, s^{-1}}$.

\begin{figure}[htp]
\includegraphics[width=520pt]{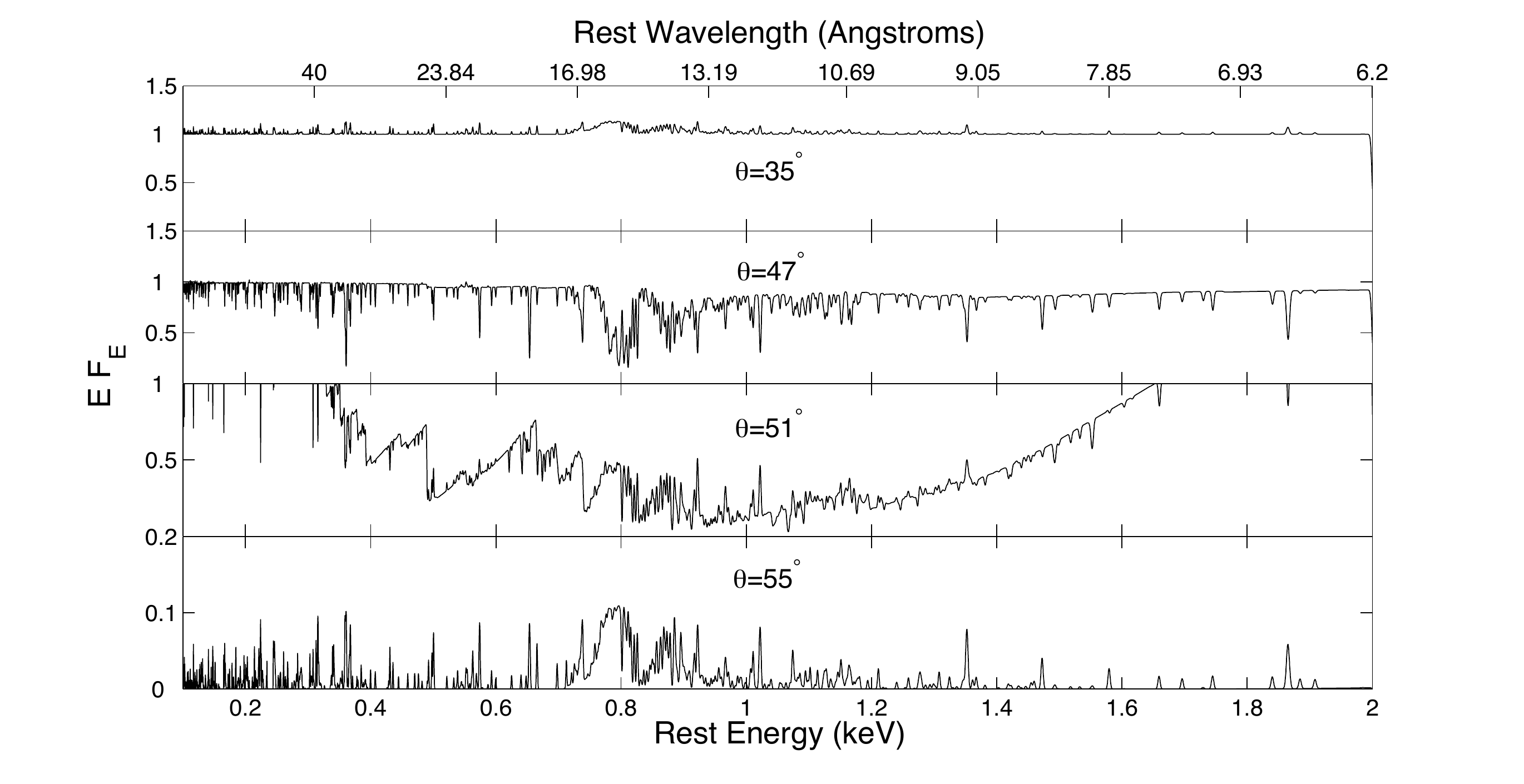}
\caption{
Model $B_3$ ($\tau_\bot^{\rm C}=40$, $R_0=0.5{\rm pc}$) X-ray spectra,
observed at $t=4$ as a function of angle. Note a different scale in 
the $y$ in the plot at the bottom; at $ \theta > 55^\circ$ the X-ray flux is strongly suppressed by the Compton thick cold matter.
}
\label{FigB3_T4}
\end{figure}

At higher energies, He-like Si K$\alpha$ line is prominent (Figure \ref{FigB3_T4}):  $ \check{\lambda}_{0}=6.6\As$ (1866 eV), $v( \check{\lambda}_{0}) \simeq160{\,\rm km\,s^{-1}}$, $v( \check{\lambda}^{+} )\simeq321{\,\rm km\,s^{-1}}$, $v( \check{\lambda}^{-} )\simeq -161{\,\rm km\,s^{-1}}$ 
At low/high inclinations, when the emission is seen on the background of either absorbed or unabsorbed continuum, an excess emission bump is seen around $16.8\As$ (740 eV); many lines which are close to $\rm O\,{VII}$, and $\rm O\,{VIII}$ are blended together contributing to this emission feature.
Observing the wind from inclinations which are ''just right'' (in our case $\sim 45^{\circ}$) allows us to see the absorption from the gas which is just on the edge of the torus; this is the case when absorption lines may show less blueshift than emission. Some emission comes from the gas which is at lower $\theta$ which is generally moving faster than the wind at high inclinations.
From
${\rm Si\,{XIII} }$ we have: $ -800< v <805 \,{\rm km\, s^{-1}}$ from $\theta=35^{\circ}$ emission, and
$ -650< v <600 \,{\rm km\, s^{-1} }$ from $\theta=45^{\circ}$ absorption.
We conclude that results for $B_{3}$ model are consistent with the general situation in which the gas is located two times closer to the BH, and the familiar funnel mechanism also helps it to escape with larger velocities: $ 50 \lesssim v  \lesssim 1000 \,{\rm km\, s^{-1}}$.

The structure of the hydrodynamical flow of this model is similar to that previously discussed:
the {\it radial} column density at $\theta=45^{\circ}$ rises from $N_{22}\simeq 0.5$ at $r=0.93\,{\rm pc}$ to 
$N_{22}=1.7$ at $r=2.9\,{\rm pc}$ up $N_{22}=38$ at $r=3.9\,{\rm pc}$; remarkably, the gain in column density occurs at the same radii as in the $B_{6}$ model. 
The line centroid, $\check{\lambda}_{0}$ as in the $B_{6}$ model, is determined by two regions of the most favorable absorption: one starting from $r\simeq 0.8\,{\rm pc}$, where $v_{r}\simeq v_{\phi} \sim 162 \,{\rm km\, s^{-1}}$ and density is $n\simeq 3\cdot 10^{3}{\rm cm^{-3}}$ and 
the other at
$r\simeq 3.56\,{\rm pc}$ where the density has a maximum value on this line-of-sight: $n\simeq 2.74\cdot 10^{5}{\rm cm^{-3}}$, and the velocity $v_{r}\simeq 37 \,{\rm km\, s^{-1}}$. 
Between these regions
at $r=1.8\,{\rm pc}$ the density decreases but the velocity increases: $v_{r}=416 \,{\rm km\, s^{-1}}$, as expected from mass flux conservation.  
As in the $B_{6}$ case, the ionization parameter, $\xi$ has a plateau, where $\xi\simeq 2000$. Then it behaves just like inverted number density, i.e. to very small values at $r\simeq3.5\,{\rm pc}$.

At high inclination the radiation of the nucleus is  highly absorbed by cold gas of the torus. There are a few emission lines seen on the background of such attenuated continuum.
Those include $\check{\lambda}_{0}\sim 13.5\As$ (921.4) eV representing a ${\rm Ne\,{\rm IX}}$ line, possibly blended with moderately ionized ${\rm Fe\,{XVII} - Fe\,{XXIV}}$ lines. 
From the broad ${\rm Ne\,{\rm X}}$ line, we obtain $\check \lambda^{+} \simeq 12.1\As$ (1024 eV),
$\check \lambda^{-}\simeq 12.2 \As$ (1020 eV), indicating $|v| \lesssim 587 \,{\rm km\, s^{-1}}$.
From
$\rm Mg\,{\rm XI}$ line, we obtain $\check \lambda^{+} \simeq 9.15\As$ (1354 eV), $\check\lambda^{-}\simeq 9.17\As$ (1352 eV), indicating $|v| \lesssim 443 \,{\rm km\, s^{-1}}$.
Maximum velocities of emission lines are reminiscent of the high velocity, low inclination outflow.
At low inclination, $\theta\simeq 10^{\circ}$, the hydrodynamic model gives $v_{r} \simeq 935 \,{\rm km\, s^{-1}}$ at $r\simeq 3.2\,{\rm pc}$ and toroidal velocity, $v_{\phi}\simeq 705 \,{\rm km\, s^{-1}}$ at $r\simeq 0.37\,{\rm pc}$;  the column density smoothly rises from $N_{22}=0.1$ at $r \simeq1\,{\rm pc}$ to $N_{22} \simeq 1$ at $r=20\,{\rm pc}$. Overall attenuation of emission lines are consistent with the picture inferred for the $B_{6}$ model attenuation (c.f. Figure \ref{FigB6_T4LineSIght}).

\subsubsection*{Results for the $A_{3}$ model} 

\begin{figure}[htp]
\includegraphics[width=520pt]{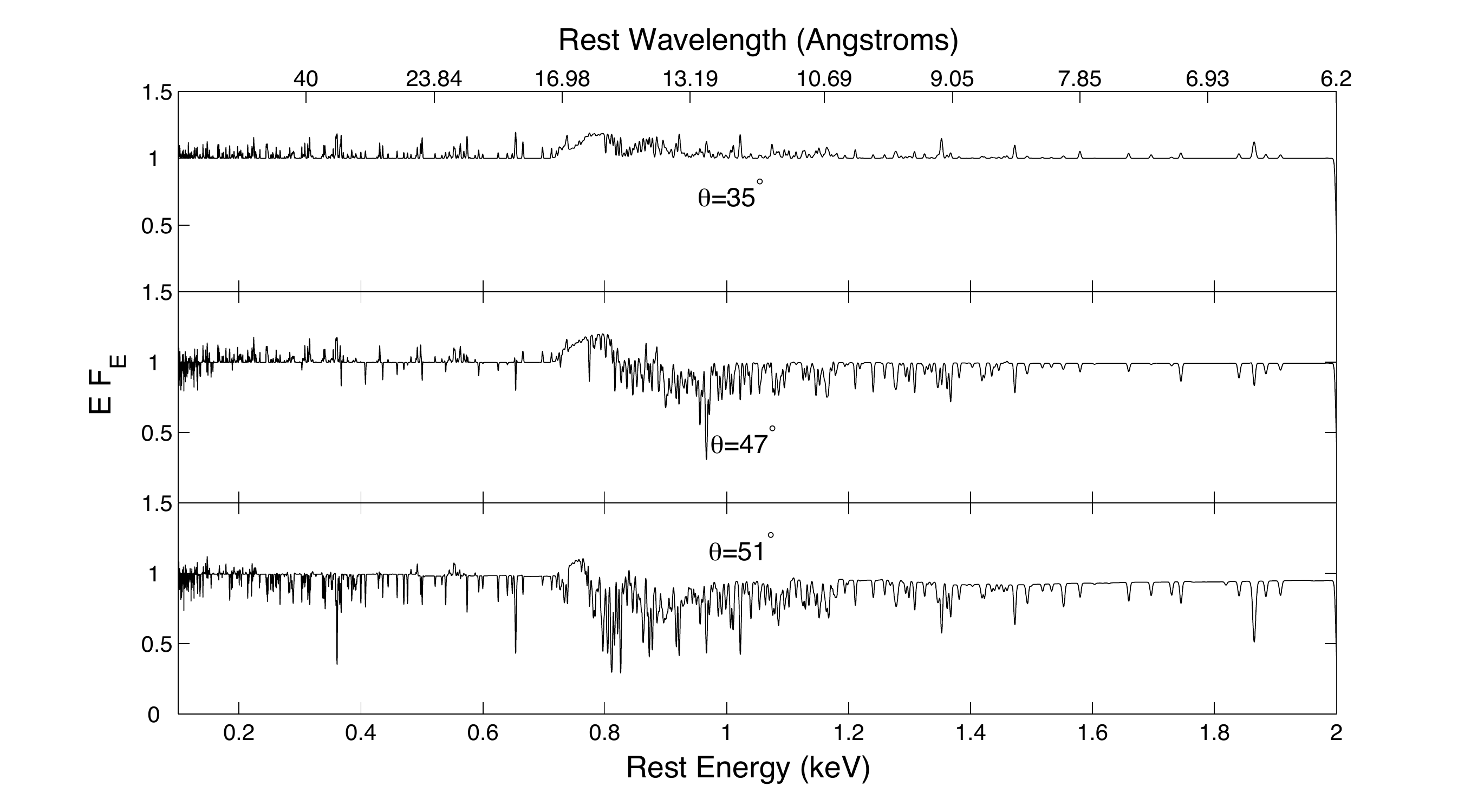}
\caption{
Model $A_3$ ($\tau_\bot^{\rm C}=1.3$, $R_0=0.5{\rm pc}$) X-ray spectra,
observed at $t=4$ as a function of angle.
At $ \theta > 55^\circ$ the X-ray flux is strongly suppressed by the Compton thick cold matter.
}
\label{FigA3_T4}
\end{figure}

Figure \ref{FigA3_T4} shows convolved emission-absorption spectra for the model $A_{3}$ ($ \tau_\bot^{\rm C}=1.3$, $R_{0}=0.5$, $\Gamma=0.5$, 
$d=2.5$) at time, $t_{0}=4$ at different inclinations.
A pure emission line spectrum is showing up at $\theta\lesssim 47^{\circ}$.
Strong lines from $\rm O\,{VII}$, $\rm O\,{VIII}$, $\rm Mg\,{XI}$, $\rm Mg\,{XII}$, $\rm Si\,{XII}$, are blended with
many other lines, in particular with those from low ionization ${\rm Fe
\,{XVII} - Fe\,{XXIV}}$. 
Inferred velocities are consistent with those known directly from the hydrodynamic model. 

Observing emission lines at low $\theta$ we see directly through the torus throat, thus having a chance to see the gas which recedes from us on the other side.
Thus, at $\theta=35^{\circ}$,
from $\rm O\,{VIII}$ K$\alpha$ line we infer $\hat \lambda^{+}= 18.9 \As$ (655.5 eV) and 
$\hat \lambda^{-}= 19\As$ (652 eV) indicating $-725<v<881\, \rm km\, s^{-1}$.
Many emission lines at higher energies have symmetrical profiles:
from $\rm Mg\,{XI}$ line we infer $\hat \lambda^{+}= 9.2 \As$ (1355 eV) and 
$\hat \lambda^{-}= 9.2 \As$ (1349 eV), indicating $-665<v<665\, \rm km\, s^{-1}$;
at higher energies, from $\rm Si\,{XIII}$ line we obtain $\hat\lambda^{+}= 6.66\As$ (1865 eV) and 
$\hat \lambda^{-}= 6.64 \As$ (1861 eV), indicating $-643<v<643\, \rm km\, s^{-1}$.
Observed from an inclination $\theta_{0}\simeq 45^{\circ}$ we look almost in the torus throat, just touching its inner evaporative skin, and these emission lines are formed by the gas within such a cone, i.e. at $0<\theta<\theta_{0}$.  
We note the likely role of the accretion disk which in real AGN is opaque to the external radiation and can screen the receding gas. Obviously, no accretion disk in our simulations is explicitly taken into account so we are not able to take this effect into account.

The lower density of models $A_{i}$ in comparison with models $B_{i}$, allow us to see more interesting profiles of individual lines.
At the critical inclination, $\theta\sim 45^{\circ}$ the spectrum resembles that obtained from $B_{3}$ or $B_{6}$ models.
However, in this case the profiles of many lines have a shape of P-Cygni and many other have emission-absorption-emission features, with absorption being considerably stronger than emission. One example of such an M-shaped profile is shown in the next section (Figure \ref{FigA3_T4_lines1}). 
The red- or blueshifted wings seen in the emission are due to a combination of rotation and the P-Cygni mechanism (in the red-ward part of the profile) and due to rotation (in the blue-ward part of the line profile). The presence of M or U shape profiles is clearly marking the problem: the importance of rotation and fast gas moving at low $r$, i.e. $v_{\phi}$, and $v_{\theta}$. Note that the appearance of such profiles from a rotating ring or disk, in the Sobolev approximation, has been emphasized by
\cite{RybickiHummer83}. We will discuss these effects in some detail further in the text.

\begin{figure}[htp]
\includegraphics[width=500pt]{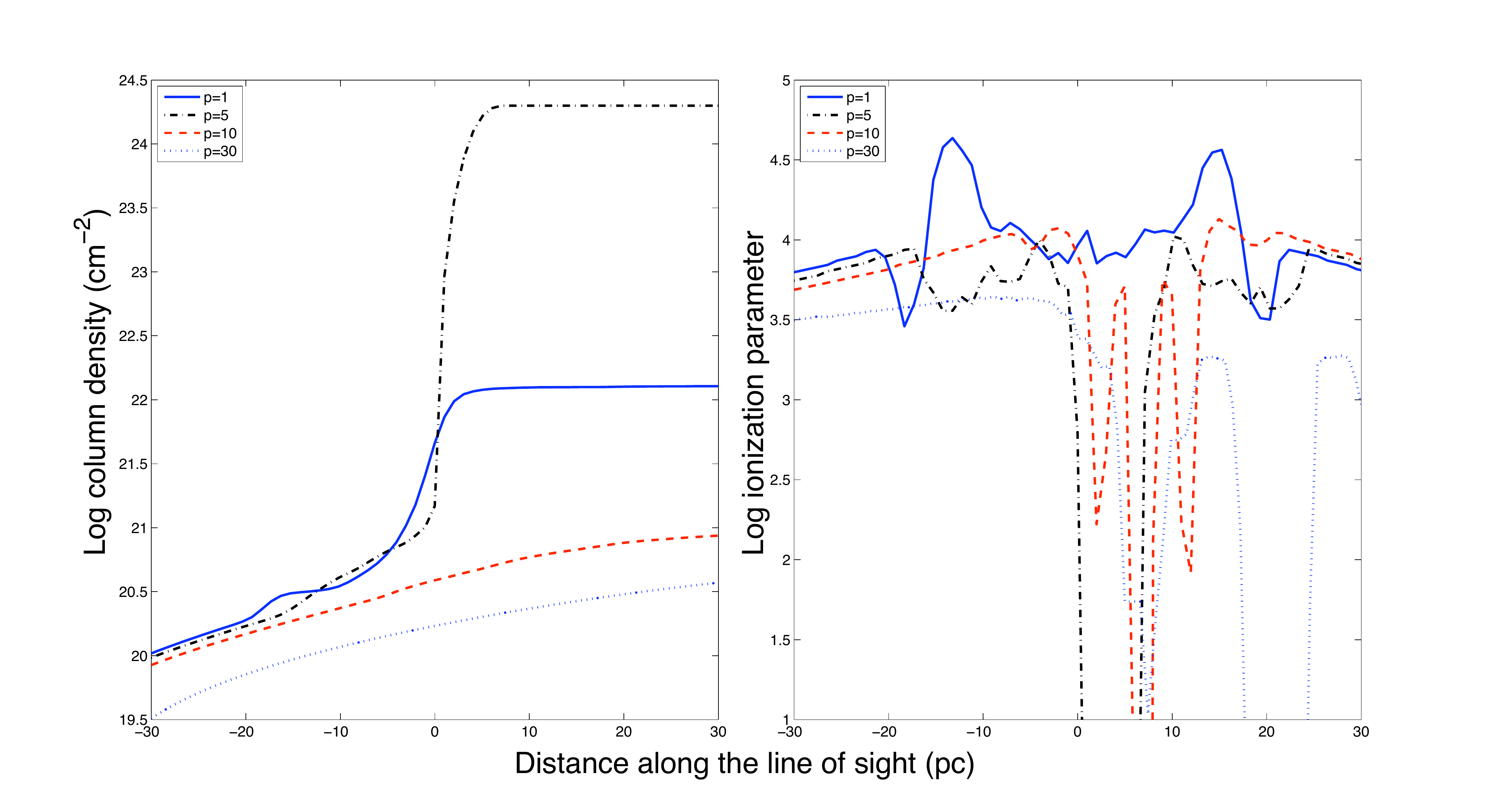}
\caption{
Model $A_3$,
observed at time $t=4$ and at inclination $\theta_{0}=45^{\circ}$ as a function of impact parameter, $p$; the angular coordinate in the plane of the sky, $\phi_{0}=45^{\circ}$. Left panel: column density; right panel: ionization parameter. Legend: each curve is calculated for the impact parameter, p, measured in parsecs.
}
\label{FigA3_T4LineSIght}
\end{figure}

Figure \ref{FigA3_T4LineSIght} shows the distribution of column density, $N(L, \theta_{0}, \phi_{0})$ and ionization parameter, $\xi(L, \theta_{0}, \phi_{0})$ along the line of sight. At $p\sim 1$, 
$L\sim 0$, the edge of the cooler slow wind or torus can be seen in the increase of N. Along the line-off-sight the ionization parameter is in the range $1 \lesssim \xi \lesssim 10^{4}$.

\section{Discussion}
Analysis of different models shows that, roughly speaking, in the middle of the torus evolution the spectrum which is most rich with spectral features is observed at inclinations near $\theta=45^{\circ}$. This holds approximately true for both types of hydrodynamical models, $A_{i}$ and $B_{i}$. At lower inclinations gas is mostly observed in emission lines while at higher inclinations, the spectrum becomes significantly absorbed by the cold gas in the torus. However some of the strong emission lines are seen atop of such absorbed continuum. 

The positions of the line centroid energies are generally consistent with what is expected from the torus X-ray exited wind paradigm, i.e. the inferred velocities are several times $V_{\rm esc}$ at 1 pc.
There is also found an influence of the gas that was evaporated off the torus and then accreted deep into the potential well; larger covering factor and high velocity allows it to block enough continuum and produce high velocity broadening of some lines. 
These high velocity signatures come from distances as small as $0.1$ pc. Another interesting effect is a blending of multiple absorption and emission lines which are seen in the un-convolved, high resolution spectrum but form a single absorption line in the convolved spectrum. Comparing high resolution and convolved spectra, we show that interpretation of such lines may in some cases lead to significant overestimation of the inferred gas velocity.

Many strong lines are observed blueshifted in absorption and at almost zero blueshift or very small in emission: $v( \hat{\lambda_{0} } )\lesssim 50\,{\rm km\,s^{-1}}$;
For example, ${\rm Mg \, XII}$ line $8.4\As$ (1472 eV) in absorption indicates $v\sim 200 \,{\rm km\, s^{-1}}$ and having in emission almost zero shift,  
$\check{\lambda}_{0} \simeq \lambda_{0} ({\rm Mg \, XII} )$ for the model $B_{3}$. However from the high resolution spectrum we see the wings of this line indicate $v\sim 600 \,{\rm km\, s^{-1}}$ (for the $B_{3}$ model)

Blending of lines is important, for example
there is a spectral feature which is observed in emission at $\theta\sim 50^{\circ}$, and in absorption at $\theta \sim 45^{\circ}$ in $B_{3}$ and $B_{6}$ models. Comparing with the un-convolved spectrum shows that this is the result of blending of many lines, specifically the H- and He-like Ne, and lines from low ionization Fe occurring in the $13.4 - 13.5\As$ (920.2-925.7 eV) range. After convolution it produces a spectral line at $13.5\As$ (921.9 eV).

\subsubsection{M shape line profiles}
Individual line profiles  obtained from some of our hydrodynamical models display multi-component features. That is, such a profile has two emission components separated by an absorption trough. Most clearly such profiles are seen in models $A_{i}$ i.e. in those models which tend to have lower density and a wider throat of the torus. U shape profiles are well known in the case of fluorescent line emission from accretion disk (see e.g. the nonrelativistic limit of \cite{Cunningham75} or \cite{GerbalPelat81}).
Using methods of escape probabilities,
\cite{RybickiHummer83} calculated a profile of a line formed within a thin rotating disk.
Contrary to the familiar reflection fluorescent line, in their case the line is formed due to resonant scattering within the disk with the resultant formation of an M - shaped line profile. 
Roughly speaking, motion of the gas distorts the M-shape profile
towards a P-Cygni shape, i.e. contributing to the red-ward emission hump. Altogether there appears to be an M-shape profile with stronger red-ward emission (Figure \ref{FigA3_T4_lines1}). 

\begin{figure}[htp]
\includegraphics[width=520pt]{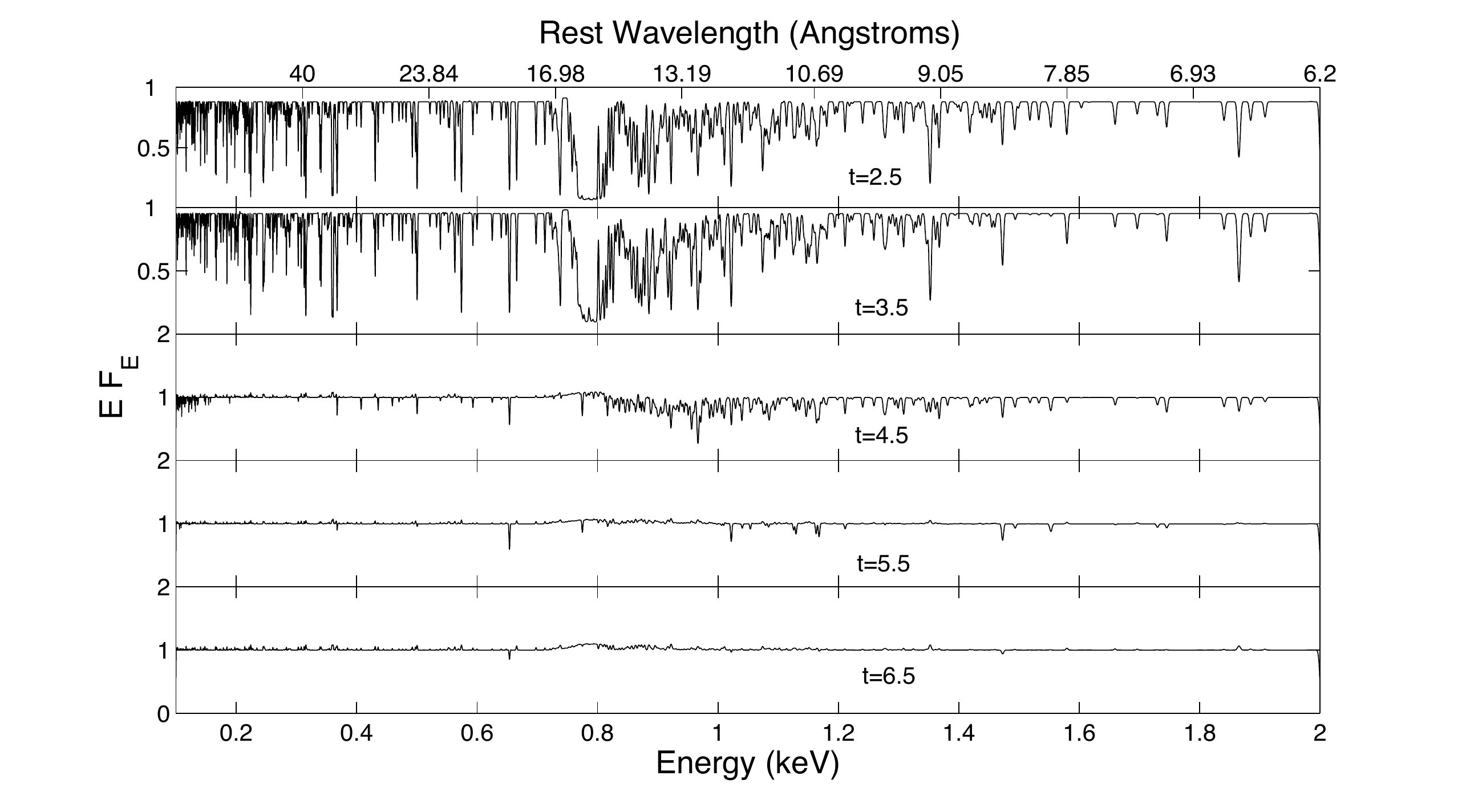}
\caption{
Model $A_6$,
observed at different evolution times at inclination $\theta_{0}=47^{\circ}$.
}
\label{A6_evolution}
\end{figure}

\subsubsection{Evolution in time}

The time evolution of model $A_{6}$ is shown in Figure \ref{A6_evolution}.

The $\rm O\,{VIII}\,{\rm L}\alpha$ is seen in absorption from $t\sim 2$ to 
$t\sim 6$. 
Starting from $t=2.5$ to $t\simeq 6$ from the high resolution spectra, we have:
$v({\check \lambda_{0}})\simeq 45 \,{\rm km\, s^{-1}}$, $v({\check \lambda^{-} })\simeq -642 \,{\rm km\, s^{-1}}$, and $v({\check \lambda^{+} })\simeq 872 \,{\rm km\, s^{-1}}$. Such characteristics remain almost unchanged until $t\sim 6$, at which the weak M-shape profile appears:  ${\check \lambda}_{0} $ remains approximately the same, while $v({\hat \lambda^{-} })\simeq -550 \,{\rm km\, s^{-1}}$,
$v({\lambda^{-} })\simeq 640 \,{\rm km\, s^{-1}}$, and $v({\hat \lambda^{+} })\simeq 642 \,{\rm km\, s^{-1}}$,
$v({\lambda^{+} })\simeq 870 \,{\rm km\, s^{-1}}$ ( $\lambda^{\pm}$ is the energy of the most red-, or blue-ward wing of the line, $\hat \lambda^{\pm}$ is related to the different emission parts of the single M-shaped line).
 
Quite different behavior is found for other lines, such as 
$\rm Si\,{XIII}$: the line is in absorption at times $t\lesssim 5$ showing weak P-Cygni profiles at $t=5.5$: $v({\lambda^{-} })\simeq -643 \,{\rm km\, s^{-1}}$, $v({\check \lambda_{0}}) = 160 \,{\rm km\, s^{-1}}$, and $v(\lambda^{+})=321 \,{\rm km\, s^{-1}}$.

During the time evolution the torus is heated and loses mass via the wind. 
Lowest density parts of it suffer most, and the torus is additionally influenced by
the combined pressure of its own wind and radiation pressure.
Altogether they work to squeeze the torus towards the equatorial plane, \citep{Dorodnitsyn08b}.

This is clearly seen in the $\rm O\,{VIII}\,{\rm L}\alpha$
line which is observed in absorption at almost all inclinations at which warm absorber is seen, and all times. Also, virial arguments suggest that the absorber in this case is located closer to the torus. 
At later times, when the torus has the shape of a thick disk, at higher inclinations we are still able to see gas close to the source, where $v_{p}\sim v_{\phi}$, and the M-shaped profile is observed.
Unfortunately, the fine structure of most lines is seen only in high resolution spectra and completely washed out by the convolution procedure.
At most times,
three dimensional modeling of our 2.5 hydrodynamical models predicts warm absorber spectrum in a relatively narrow range of inclinations: $\Delta\theta\lesssim 10^{\circ}$.
 
\subsubsection{Fe line region}
\begin{figure}[htp]
\includegraphics[width=520pt]{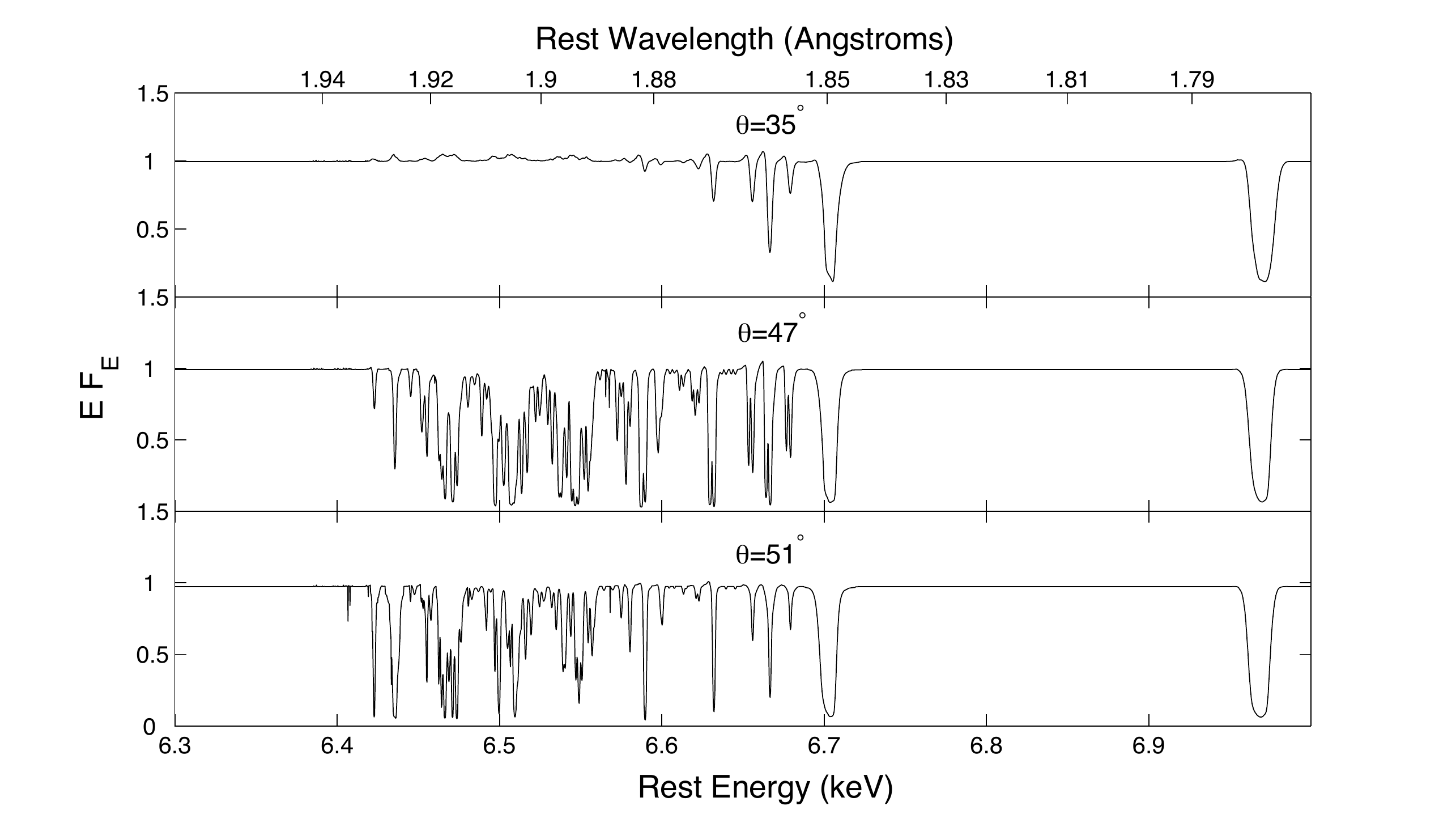}
\caption{
Model $A_3$, observed at $t=4$ a different inclinations. Notice prominent absorption lines from ${\rm Fe\,{XXV}}\, 1.85 \As$ (6.701 keV) and 
${\rm Fe\,{XXVI}}\, 1.78 \As$ (6.96 keV).
}
\label{FigA3_T4_Fe1}
\end{figure}

We have calculated spectra from a torus wind in the 6.4 keV region.
The results for the $A_{3}$ model are shown in Figure \ref{FigA3_T4_Fe1} as a function of inclination, $\theta$. 
Since the fluorescent ${\rm Fe\,K}$ $\alpha$ line cannot be produced by resonance scattering, we do not have it in our spectra.

There are two prominent absorption features from resonant absorption in 
${\rm Fe\,{XXV}}$ $1.85\As$ (6701 eV) and ${\rm Fe\,{XXVI}}\, 1.78\As$ (6960 eV). Both lines have $v({\check \lambda_{0}})\simeq 100
 \,{\rm km\, s^{-1}}$. Blueshifts are consistent with that expected from the flow becoming slower at higher inclinations.
The centroid energy is located at moderate blueshifts: for ${\rm Fe\,{XXVI}}$, at $\theta=35^{\circ}$, $v( \check{\lambda}_{0})=170 \,{\rm km\, s^{-1}}$, and $v(\check \lambda^{+})=688 \,{\rm km\, s^{-1}}$, becoming at
$\theta=45^{\circ}$, $v( \check{\lambda}_{0})\lesssim 100 \,{\rm km\, s^{-1}}$, and $v(\check \lambda^{+})=688 \,{\rm km\, s^{-1}}$. 

Analogously, from the $\rm Fe\,{ XXV}$ line we have: 
at $\theta=35^{\circ}$, $v( \check{\lambda}_{0})=179 \,{\rm km\, s^{-1}}$, and $v(\check{\lambda}^{+})=671 \,{\rm km\, s^{-1}}$, becoming at
$\theta=45^{\circ}$, $v( \check{\lambda}_{0})=134 \,{\rm km\, s^{-1}}$, and $v(\check \lambda^{+})=573 \,{\rm km\, s^{-1}}$.

\begin{figure}[htp]
\includegraphics[width=420pt]{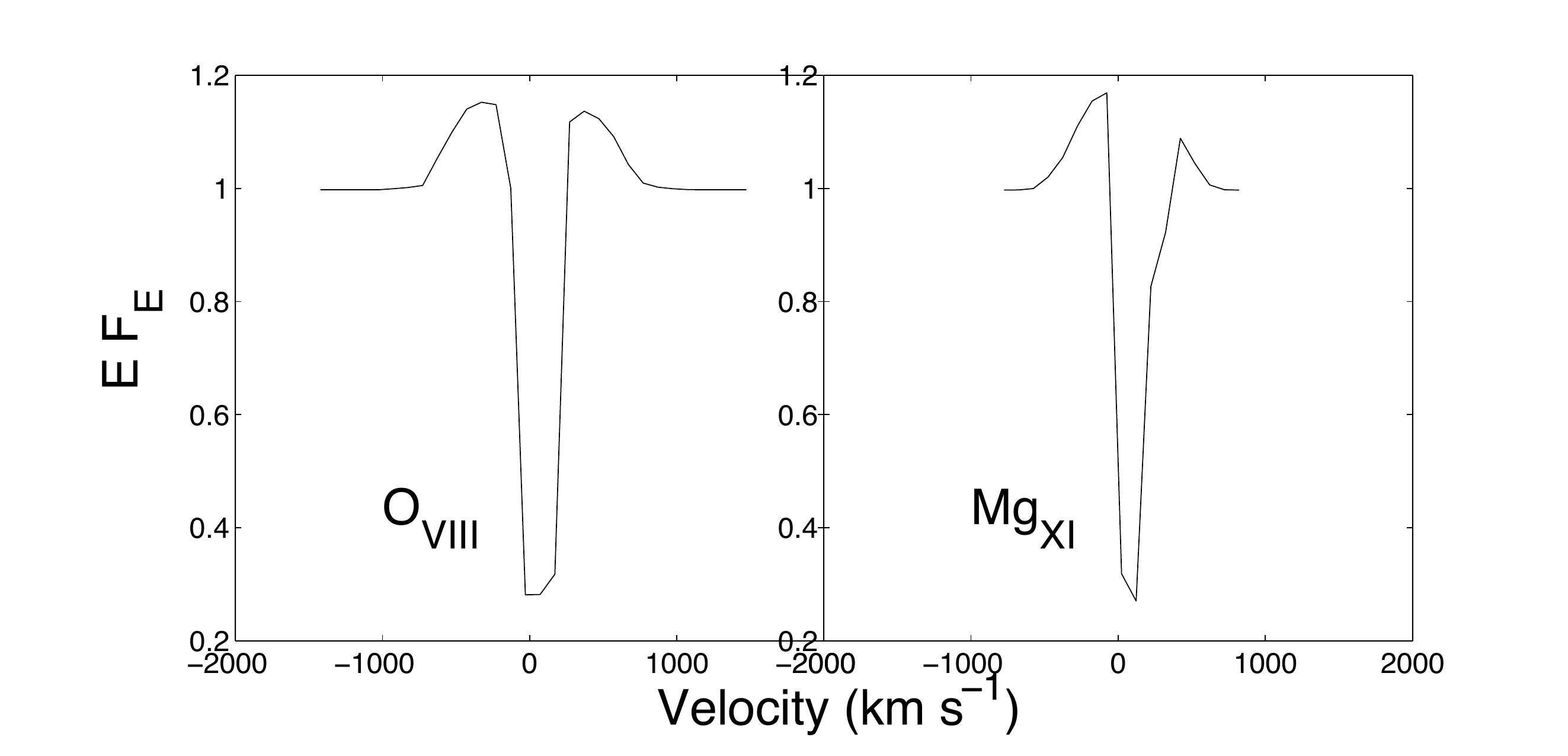}
\caption{
Model $A_3$, observed at $t=4$ a $\theta=45^{\circ}$. Profiles of individual lines of ${\rm O\,{VII}}$ (left) and ${\rm Mg\,{XI}}$ (right).
Blueshifts correspond to positive velocities, redshifts to negative 
}
\label{FigA3_T4_lines1}
\end{figure}

The results for the ${\rm O\,{VII}}$ L$\alpha$ line and for ${\rm Mg\,{XI}}$ lines are shown in Figure \ref{FigA3_T4_lines1}.
Thus, from
$\rm O\,{VIII}$ L$\alpha$ line we obtain $\hat \lambda^{-}_{\rm max}= 19\As$ (652.9 eV),
$\hat \lambda^{+}_{\rm max}=18.95\As$ (654.4 eV) where  $\hat \lambda^{-}_{\rm max/min}$ is the energy at the maximum/minimum of the corresponding emission component. From this we derive $v(\hat \lambda^{-}_{\rm max})=-312\, \rm km\, s^{-1}$ and $v(\hat \lambda^{+}_{\rm max})=376\, \rm km\, s^{-1}$. Analogously,
we obtain $v(\hat \lambda^{-})=-725\, \rm km\, s^{-1}$ and $v(\hat \lambda^{+})=789\, \rm km\, s^{-1}$. 

Performing the same analysis, for example, for the $\rm Mg\,{XI}$ line we obtain:
$v(\hat \lambda^{-}_{\rm max})\simeq 0\, \rm km\, s^{-1}$ and $v(\hat \lambda^{+}_{\rm max})=443\, \rm km\, s^{-1}$, and for the maximum value of the velocity
we obtain $v(\hat \lambda^{-})=-665\, \rm km\, s^{-1}$ and $v(\hat \lambda^{+})=887\, \rm km\, s^{-1}$. 
Thus it appears for this model the Fe L$\alpha$ are moving lightly more slowly than the ${\rm O\,VIII}$ or ${\rm Mg\,XI}$ lines.

\begin{figure}[htp]
\includegraphics[height = 200pt]{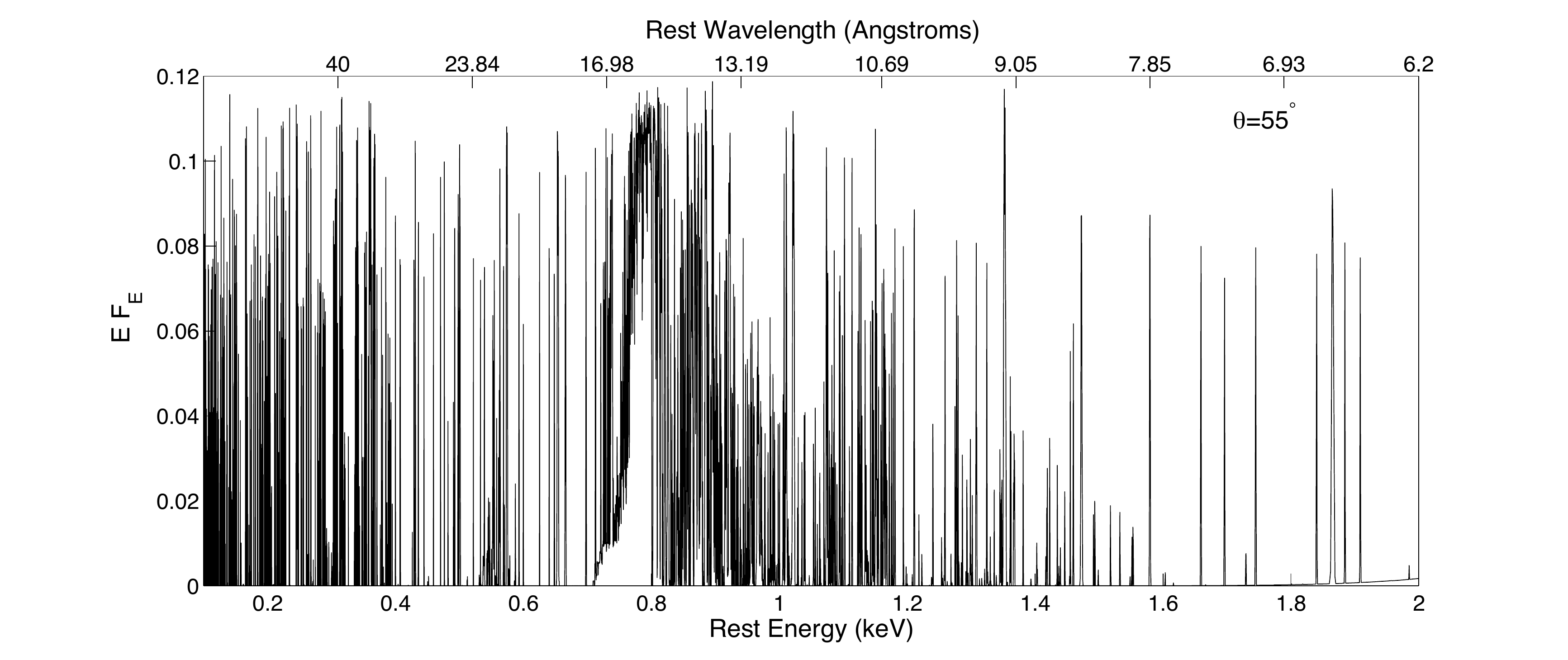}
\caption{
Model $B_3$, observed at $t=4$ a $\theta=55^{\circ}$. Unconvolved spectrum is shown. Notice, blending of many lines due to O and low ionization Fe, near 0.8 keV.
}
\label{FigB3_th55_T4}
\end{figure}

\subsubsection{Spectrum at high inclinations}
At inclinations, $\theta\gtrsim 50-60^{\circ}$, continuum radiation of the nucleus is heavily obscured, and only radiation scattered in lines is seen. Figure \ref{FigB3_th55_T4} shows the unconvolved spectrum from B3 model at t=4, $\theta=55^{\circ}$ revealing numerous emission features. Comparing with the prototypical NGC 1068, Syfert 2 spectrum of \cite{Kinkhabwala02}, we can see that many lines observed in emission in AGN spectrum can be identified in our simulations. However in general, the method adopted in this paper tends to overestimate the radiation flux scattered in lines.

The apparent emission bump seen in the Figure \ref{FigB3_th55_T4} at $\sim
16 - 16 \As$ is significantly more prominent than in observations of \cite{Kinkhabwala02}.  Individual lines which contribute to this hump such as lines from Ne IX, 
Fe XIX, Fe XIII, O III, and O VII and moderately ionized iron, Fe XVII - Fe XXIV are observed in NGC 1068 and are easily identifiable in our simulations.  Notice that in our method we treat all lines as resonant.  This simplification does not lead to any significant errors when calculating the absorption spectrum, although in emission more radiation is scattered in lines as it would be if the source function (\ref{SourceFunction}) would include ''thermal'' term. In the present studies we thus overestimate the scattered flux, and a good example of it is the prevalence of the emission about 15$\As$ compared to real AGNs. 

At inclinations of $70-80^{\circ}$ the spectra of B models become completely black. As for A models, for example, at inclinations as high as $80^{\circ}$ at t=4.5 the $A_{3}$ has a spectrum, which is very similar to that shown in the Figure \ref{FigB3_th55_T4}.
Overall emission is on the 10\% level of that of the continuum. Notice an emission excess near $15.5\As$ (0.8 keV), which we have discussed in the context of properties of spectra from A and B models.
The spectrum contains numerous features from $\rm O$, $\rm Mg$, $\rm Si$, Ne, and  low ionization $\rm Fe$. The existence of such spectrum suggests a continuous distribution of ionization parameter, $\xi$ over the scattering region in qualitative correspondence with observations of Seyfert 2 galaxies \citep{Kinkhabwala02,Brinkman02}.

\section{Conclusions}
We have carried out 3D radiation transfer simulations of the absorption-emission spectra produced by the X-ray excited wind from the obscuring torus in AGN.
As a previous stage of our research we performed time-dependent 2.5D numerical simulations of wind which originates from 
cold, rotationally supported torus whose inner throat is exposed to a strong ionizing radiation of the BH accretion disk
\citep{Dorodnitsyn08a,Dorodnitsyn08b}. Previous results include purely absorption spectra from such a flow, and suggest that this model is promising in establishing cold tori as a major source of the warm absorber flows observed in many AGN. 

Motivated by the fact that
in case of an extended wind a 1D transmission approximation for the radiation transfer in lines is inadequate, we relax this approximation
by performing 3D simulations of the radiation transfer in X-ray lines in a moving gas of the torus wind. The distribution of the gas is obtained from the hydrodynamical modeling which is used as an input to the radiation transfer calculations carried out making use of a 3D generalization of the Castor escape probability formalism \citep{RybickiHummer78,RybickiHummer83}.

The richest warm absorber spectra are observed within $\Delta\theta\simeq 10^{\circ} - 15^{\circ}$, and the inferred velocity of the gas decreases as inclination increases. 
Synthetic spectra predicted by our hydrodynamical models are generally consistent with observations of warm absorbers.  They contain many 
strong absorption as well as emission lines from $\rm O$, $\rm Mg$, $\rm Si$, Ne, and  $\rm Fe$.
Numerous features 
found in our synthetic spectra near $13.7 - 13.9 \As$ (892 - 905 eV), $9.3 - 9.4\As$ (1319 - 1333 eV), and $6.7 - 6.9 \As$ (1797 - 1831 eV) are likely due to
absorption from high ionization Ne, Mg and Si, respectively.  
Many lines found in our synthetic spectra
are blended, in particular with those from  ${\rm Fe\,{XVII} - Fe\,{XXIV}}$.
Velocities of this medium ionization component which are inferred directly from spectral lines shift are consistent with those derived directly from hydrodynamical models. 
Some of the individual line profiles demonstrate P-Cygni profiles and others have weak M-shaped profiles, with the red-ward wing being usually stronger than the blue-ward wing. 
Such a profile is the result of complicated flow geometry at smaller radii: 
i.e. a combination of P-Cygni profile and the importance of the azimuthal, $v_{\phi}$ and poloidal, $v_{\theta}$ velocities at smaller radii.

If a warm absorber flow is thermally driven or ablated off of a cold or nearly-neutral surface, then the observed low
ionization component provides evidence for this origin. 
This is confirmed by our numerical simulations: at the angles at which the line of site goes through the X-ray evaporated ''skin'' of the torus we see such a stripped gas.
Observations also suggest the simultaneous presence of 
gas at a low ionization state, log $\xi \simeq 0$ and a higher ionization component with log $\xi \simeq 100$.
Note also some claims of observational evidence
ruling out the existence of gas at intermediate ionization parameters
 \citep{McKernan07, Holc07} for some objects. 
Gas at ionization parameters outside the range 0$\leq$log($\xi)\leq$3 is harder to detect 
unambiguously from observations. 

Torus winds naturally possesses features which are necessary for the formation
of the flow with characteristics typical for warm absorbers.
Our calculations show that the angular distribution of $\xi$ is strongly angle-dependent, reflecting the 2D nature of the flow.

Our solutions also show that $\xi$ can vary dramatically along the line-of-sight.
 In absorption, at such inclinations at which the warm absorber is observed, the line-of-sight first touches the stripped cooler gas and then penetrates the torus funnel, where hotter gas is located. Emission line photons are collected from a larger volume, coming from separated places 
in the outflowing gas, those which are not significantly obscured by the cold gas
of the torus. 
In such flow different photons in a single line may come from regions quite separated in space (multiple resonances) making a single zone approximation for $\xi$ inadequate.

Our numerical methods are not capable of capturing some physics which is probably of importance here, most notably the co-existance of multiple phases in the outflowing plasma on level smaller than a typical size of our numerical cell. 
Bearing in mind all these limitations, we note that in our hydrodynamic calculations we don't see specific strong bi-modality in the distribution of $\xi$.

Our models divide in two broad categories: more dense, having initial Compton optical depth of the torus, $ \tau_\bot^{\rm C}= {\rm few}\cdot10$, and less dense, with $ \tau_\bot^{\rm C}\simeq1$. Additionally, we consider cases in which we put the initial torus, at either 1pc, or closer, 0.5pc from the BH. The X-ray heated gas is accelerated by the thermal mechanism within the throat of the torus.  As already established 
\citep{Dorodnitsyn08b}
such a funnel mechanism helps the gas to accelerate to velocities, $V_{\rm max}$ of the order of several escape velocities, $V_{\rm esc}$(1pc). This velocity, $V_{\rm max}$ is not necessarily equal to the terminal velocity, $V^{\infty}$, due to the properties of the funnel flow.

Results from our new method generally agree with the pure transmission calculations of \cite{Dorodnitsyn08b}. That is, warm absorbers are observed roughly in a $10^{\circ}$ range.
In general, the velocity corresponding to the centroid energy of an absorption line, is smaller than $300\, \rm km\, s^{-1}$. Most of such absorption is produced in a hot X-ray evaporated skin of a torus. At such distances, the finite size of an X-ray source is not important.
However, the blue-ward part of the absorption line in many cases moves as fast as $1000\, \rm km\, s^{-1}$, meaning the absorber is at least partially located at smaller radius. At such small radii the core can no longer be treated as a point source.  Rather it subtends a finite solid angle 
which finally  translates to a broader angular pattern at which the absorption spectrum is seen from infinity. Such an effect is revealed from individual line profiles. The most rapidly moving gas ($v \gtrsim 500-800$ km$s^{-1}$) sees such core from close proximity, and that influences the maximum blue- and redshift of photons within the line profile. 

This suggests that the torus wind may extend inward to the region where considered for accretion disk winds.  
As it could be impossible to tell between the torus wind and an accretion disk wind at such small distances from a BH,
this result implies an accretion disk wind should be considered as a possibly important ingredient of modeling of warm absorber flows.

\hbox{}
This research was supported by an appointment to the NASA Postdoctoral Program at the NASA Goddard Space Flight Center, administered by Oak Ridge Associated Universities through a contract with NASA, and by grants from the NASA Astrophysics Theory Program 05-ATP05-18.

\end{document}